\documentclass[twocolumn,tighten]{aastex7}

\usepackage{amsmath,amstext,amssymb}
\usepackage{graphicx}

\begin{document}

\title{A Hybrid Algorithm for Drift-Kinetic Particle Dynamics within General Relativistic Magnetohydrodynamics Simulations of Black Holes Accretion Flows}

\author[0009-0004-8116-3123]{Tyler Trent}
\affiliation{School of Physics, Georgia Institute of Technology, 837 State St NW, Atlanta, GA 30332, USA}
\affiliation{Departments of Astronomy and Physics, University of Arizona, 933 N. Cherry Ave., Tucson, AZ 85721, USA}
\email{ttrent@arizona.edu}

\author[0000-0003-1035-3240]{Dimitrios Psaltis}
\email{dpsaltis3@gatech.edu}
\affiliation{School of Physics, Georgia Institute of Technology, 837 State St NW, Atlanta, GA 30332, USA}

\author[0000-0003-4413-1523]{Feryal \"Ozel}
\email{feryal.ozel@gatech.edu}
\affiliation{School of Physics, Georgia Institute of Technology, 837 State St NW, Atlanta, GA 30332, USA}

\begin{abstract}
Astrophysical plasmas in relativistic spacetimes, such as black hole accretion flows, are often weakly collisional and require kinetic modeling to capture non-local transport and particle acceleration. However, the extreme scale separation between microscopic and macroscopic processes limits the feasibility of fully kinetic simulations. A covariant guiding center formalism has recently been derived to address this challenge in curved spacetimes. We present a new hybrid numerical algorithm based on this formalism, which evolves the trajectories of charged particles over macroscopic timescales in GRMHD backgrounds. To address numerical instabilities in the equations of motion, we develop a semi-implicit integrator that ensures stable evolution in strong-field environments. We apply our method to GRMHD simulations of black hole accretion flows, demonstrating its accuracy and efficiency across a range of physical conditions.
\end{abstract}

\section{Introduction}
\label{sec:intro}

Astrophysical plasmas are often the source of light that enables our astronomical observations. Understanding their fundamental behavior is, therefore, crucial for interpreting observed phenomena. In many settings, including the solar wind, the intracluster medium, and accretion flows around black holes, plasmas are either collisionless or weakly collisional, owing to the low particle density. These conditions necessitate modeling via kinetic theory, which solves self consistently for the motion of the plasma particles and the fields they produce. 

Standard kinetic approaches are constrained to resolving the gyromotion and plasma skin depth, i.e., scales that are vastly smaller than those of the astrophysical systems under study. For example, the gyroradius of a typical electron in the inner accretion flow of the black hole Sagittarius~A$^*$ (Sgr~A$^*$) is at least $10^8$ times smaller than the Schwarzschild radius \citep{Trent2023ASpacetimes}. This disparity severely limits the dynamic range achievable in kinetic simulations, confining most studies to microscopic regions, even when employing state-of-the-art massively parallel numerical algorithms \citep[see, e.g.,][]{Sironi2009PARTICLEOBLIQUITY, Sironi2014RELATIVISTICPARTICLES, Werner2018Non-thermalReconnection, Kunz2014FirehosePlasma, Zhdankin2023SynchrotronInstability, Hakobyan2023RadiativeM87}

Because of the significant scale separation, global studies of astrophysical systems typically employ a magnetized fluid model, solving the magnetohydrodynamic (MHD) equations, or their general relativistic formulation (GRMHD) \citep[see, e.g.,][]{, DeVilliers2003GlobalTori,Gammie2003HARM:Magnetohydrodynamics}. This approach has significantly advanced our understanding of the macroscopic dynamics of plasmas. However, it fails to account for the non-local transport associated with the large mean-free paths of the charged particles. It also neglects particle acceleration mechanisms that produce populations of non-thermal electrons (see, e.g., \citealt{Rowan2017ElectronReconnection, Ball2018ElectronMagnetization, Ball2019TheReconnection, Sironi2021Reconnection-drivenFlows}).

In a third approach, it is possible to reduce the separation of scales in a kinetic model by using a guiding center formalism. This approach decomposes the trajectory of a charged particle into the gyromotion around the magnetic field, which can be integrated out analytically, and the motion of the so-called guiding center  \citep{Vandervoort1960TheField, Northrop1963TheParticles, Cary2009HamiltonianMotion}. The latter typically evolves at timescales that are much longer than the gyroperiod, allowing for macroscopic integration step sizes. This method has facilitated studies on the motion of charged particles in a variety of astrophysical systems, particularly in the context of the solar corona \citep{Gordovskyy2010ParticleEvent, Gordovskyy2010PARTICLEFLARES, Gordovskyy2011PARTICLELOOP, Gordovskyy2023ParticleField, Ripperda2017ReconnectionPlasmas, Ripperda2017Reconnection2.5D, Threlfall2018FlareRopes,  Bacchini2024ParticleLoops}. However, most of these studies have been limited to flat spacetimes, with \cite{Bacchini2020AMagnetospheres} presenting a limited exploration in black-hole spacetimes, in which only one of the relevant drift velocities of the guiding center has been taken into account.

In recent work, we extended the guiding center formalism to curved spacetimes \citep{Trent2023ASpacetimes,Trent2024CovariantSpacetimes}. We derived a fully covariant set of guiding center equations of motion that incorporate implicitly all known drift velocities of the guiding center, including the covariant extension of the gravitational drift. In the traditional approach, one uses these equations of motion to derive expression for the drift velocities, which can then be integrated to calculate the trajectories of the charges. However, this is impractical with our covariant formalism, because of the non-linearity of the general relativistic equations. Instead, we opt to integrate numerically the acceleration equations for the charges. 

In this paper, we introduce a novel, hybrid, semi-implicit numerical algorithm for integrating the covariant guiding center equations in a GRMHD background. This semi-implicit scheme is designed to overcome a numerical instability that causes explicit integrators to fail when the timestep exceeds a gyroperiod. To ensure stability and accuracy, the algorithm treats the unstable terms implicitly while retaining an explicit treatment of the stable terms, allowing for analytical forward stepping. It also incorporates high-order interpolation of the electromagnetic field tensor derived from GRMHD simulations.

In \S\ref{sec: methods}, we introduce the covariant guiding center equations of motion and the semi-implicit scheme we have developed to address the numerical instability in the equations. In \S\ref{sec:GRMHD background}, we present the integration with a GRMHD background. In \S\ref{sec: numerical characteristics}, we explore the performance of the numerical algorithm by studying the dependence of the results on magnetic field strength, integration technique, and grid resolution. Finally, in \S\ref{sec:conclusion}, we present a discussion of our findings and explore potential applications.

\section{Integrating the Covariant Guiding Center Equations}
\label{sec: methods}

In this section, we first introduce the covariant guiding center equations of motion, which are the focus of our numerical integration efforts. Subsequently, we address the issue of the numerical instability inherent in the equations and introduce the semi-implicit numerical integration scheme we developed to effectively integrate them. Throughout this paper we use units such that $G=c=1$, where $G$ is the gravitational constant and $c$ is the speed of light.
 
\subsection{General Relativistic Guiding Center Equations of Motion}
\label{subsec: GC equations}

The fully covariant guiding center equations of motion, as derived in \citet{Trent2024CovariantSpacetimes}, approximate the drift motion to first order in gyroradius. These equations encompass all traditional drifts, including the general relativistic gravitational drift, and their covariance ensures applicability across general spacetimes and arbitrary coordinate systems. 

The gyroradius, $\rho$, and gyrofrequency, $\omega$, of a charged particle are defined as \citep{Trent2024CovariantSpacetimes}
\begin{equation}
    \label{eqn: gyroradius}
    \rho \equiv \frac{u^\alpha\sigma^*_\alpha }{i\omega},
\end{equation}
\begin{equation}
    \label{eqn: gyrofrequency}
    \omega  \equiv \frac{q}{2m}\biggl\{F^{\alpha\beta}F_{\alpha\beta}+\Bigl[(F^{\alpha\beta}F_{\alpha\beta})^2+(F^{\alpha\beta}\mathord\star F_{\alpha\beta})^2\Bigr]^{1/2}\biggr\}^{1/2}\;.
\end{equation}
Here, $u^\alpha$ is the four-velocity of the charged particle, $q/m$ is the charge-to-mass ratio, $F_{\alpha\beta}$ is the electromagnetic field tensor,$\mathord\star F_{\alpha\beta}$ is its dual, $\sigma$ is the eigenvector of the electromagnetic field tensor that corresponds to the positive gyrofrequency, and $\sigma^*$ is its complex conjugate.

The guiding center approximation assumes that the electromagnetic field is slowly varying in both space and time compared to the gyroradius and gyrofrequency, respectively. We express these assumptions in terms of the dimensionless ratios
\begin{align}
\label{eqn: gc assumption 1}
    \Psi_1\equiv \frac{1}{\rho}\frac{\omega}{\text{Max}\Bigl|(e^{i\varphi}\sigma^\nu+e^{-i\varphi}\sigma^{\nu *})\partial_\nu\frac{q}{m} F^\alpha_{~\beta}\Bigr|}\sim \frac{B}{\rho (\partial B/\partial x)_\perp},
\end{align}
and
\begin{align}
\label{eqn: gc assumption 2}
    \Psi_2\equiv \frac{\omega}{2\pi} \frac{\omega}{\text{Max}\Bigl|\mathcal{U}^\nu\partial_\nu\frac{q}{m} F^\alpha_{~\beta}\Bigr|}\sim \frac{B}{(2\pi{\cal U}/\omega) (\partial B/\partial x)_\parallel},
\end{align}
where $\mathcal{U}$ is the four-velocity of the guiding center.

The first quantity, $\Psi_1$, is proportional to the magnitude of the electromagnetic field, because of the covariant gyrofrequency in the numerator, and inversely proportional to the change of the electromagnetic field in the plane of gyration across a gyroradius, estimated using a directional derivative of the electromagnetic field tensor in the direction of eigenvectors that corresponds to the gyrofrequency. The second quantity, $\Psi_2$, is also proportional to the magnitude of the electromagnetic field but inversely proportional to the change in the electromagnetic field over a gyroperiod along the direction of the guiding center motion. Together, these two ratios provide a measure of the validity of the guiding center assumptions, with larger the ratio corresponding to increased validity.

By extending the guiding center formalism to general relativistic spacetimes, we introduce one more assumption, i.e., that the effects on the particle trajectory due to gravity must be weaker than the effects due to the electromagnetic field, i.e., that
\begin{equation}
    \label{eqn: gravity gc assumption}
    \biggl|\Gamma^\alpha_{\beta\nu}u^\beta u^\nu\biggr| \ll \biggl|\frac{q}{m}F^\alpha_{~\beta}u^\beta\biggr|\;,
\end{equation}
where $\Gamma^\alpha_{\beta\nu}$ are the Christoffel symbols.

When all the guiding center assumptions are satisfied, we can decompose the 4-position of the particle, $x^\alpha$, into its position along the gyromotion and the 4-position of the guiding center, $\chi^\alpha$, as
\begin{align}
\label{eqn:GC four-pos ansatz}
    x^\alpha =  \sqrt{\frac{\mu}{\omega}}e^{i(\omega\tau+\varphi)}\sigma^\alpha 
                +\sqrt{\frac{\mu}{\omega}}e^{-i(\omega\tau+\varphi)}\sigma^{\alpha*}
                +\chi^\alpha\;.
\end{align}
Here, $\mu$ is the covariant magnetic moment of the particle, which (to zeroth order) is given by
\begin{equation}
\label{eqn: magnetic moment}
    \mu = \rho^2 \omega,
\end{equation}
and is typically considered a conserved quantity in the guiding center approach \citep{Burby2020GeneralSystems, Stephens2017OnConservation,Brizard2022OnField}. Note that this definition of the magnetic moment differs from others by the charge of the particle.

Utilizing equation~\eqref{eqn:GC four-pos ansatz}, \citet{Trent2024CovariantSpacetimes} derived an acceleration equation for the guiding center by analytically integrating out the gyromotion, which we write in fully covariant form as
\begin{equation}
    \label{eqn:GC EOM}
    \frac{d\,\mathcal{U}^\alpha}{d\tau}=-\Gamma^\alpha_{\beta\nu}\mathcal{U}^\beta\mathcal{U}^\nu
    +\frac{q}{m}F^\alpha_{~\beta}\mathcal{U}^\beta
    -\mu\nabla^\alpha\omega\;.
\end{equation} 
Here $\tau$ is the proper time.

To solve equation~\eqref{eqn:GC EOM}, it is necessary to specify the initial position and velocity of the guiding center. Typically, the initial position and velocity of the particle are known and could be used to determine the initial conditions of the guiding center by employing the original ansatz in equation~\eqref{eqn:GC four-pos ansatz}. However, because the four-position is not a true four-vector, we cannot, in general, solve for the individual components of the initial guiding center position from equation~\eqref{eqn:GC four-pos ansatz} by simple arithmetic, i.e., by isolating the guiding center position on one side of the equality\footnote{For example, in a flat spacetime with Cartesian coordinates, position vectors can be combined arithmetically, but this is not feasible in spherical coordinates.}. Instead, we take the initial position of the guiding center to be the same as the initial position of the particle. In a manner that is consistent with the guiding center methodology, the resulting error in the initial position of the guiding center is equal to at most one gyroradius.

The four-velocity, on the other hand, is a true four-vector, allowing us to derive explicitly the initial four-velocity of the guiding center. We start by differentiating equation~\ref{eqn:GC four-pos ansatz} for the position with respect to proper time and disregard higher-order terms in the four-velocity, such as those that arise from derivatives of the gyrofrequency. We then simplify this expression using the definition of the magnetic moment and gyroradius (eqs.~\ref{eqn: magnetic moment} and \ref{eqn: gyroradius}), which yields
\begin{align}
\label{eqn: initial GC velocity}
    \mathcal{U}^\alpha(\tau=0) = u^\alpha - \sigma^\alpha \sigma^*_\beta u^\beta - \sigma^{\alpha*}\sigma_\beta u^\beta.
\end{align}

The motion of the guiding center does not strictly adhere to the traditional conservation laws for massive particles, i.e., $\mathcal{U}^\alpha \mathcal{U}_\alpha \neq-1$. However, there exists an analogous norm for the guiding center velocity that incorporates a term related to the first adiabatic invariant of a charged particle, namely the covariant magnetic moment of the gyromotion. The analogous conserved quantity for the guiding center is \citep{Trent2024CovariantSpacetimes}:
\begin{equation}
    \label{eqn:conserved quantity}
    \mathcal{U}^\alpha\mathcal{U}_\alpha+2\mu\,\omega=-1.
\end{equation}
We use this conservation law to update the time component of the guiding center four-velocity in our numerical integration scheme, ensuring that this equality is maintained. 

\subsection{Stability Analysis}
\label{subsec: numerical instability}

The second term on the right-hand side of the covariant guiding center equation of motion (eq.~\ref{eqn:GC EOM}) introduces a numerical instability in standard explicit integration schemes. To clarify the origin of this instability, we analyzed a simplified scenario: flat spacetime, non-relativistic velocities, and constant, mutually perpendicular electric and magnetic fields. In this case, the guiding center equation becomes
\begin{equation}
    \label{eqn: numerical instability 1}
    \dot{\boldsymbol{\mathcal{U}}} = \frac{q}{m}\mathbf{E} + \frac{q}{m}\boldsymbol{\mathcal{U}}\times\mathbf{B}\;.
\end{equation}

This equation has an analytic solution for the drift velocity perpendicular to the magnetic field \citep{Northrop1963TheParticles},
\begin{equation}
    \boldsymbol{\mathcal{U}_0} = (\mathbf{E}\times\mathbf{B})/B^2\;.
\end{equation} 
Introducing a small perturbation in the velocity, $\boldsymbol{\mathcal{U}}=\boldsymbol{\mathcal{U}_0}+\boldsymbol{\epsilon}$, leads to
\begin{align}
    \label{eqn: numerical instability 2}
    \frac{d}{dt}\biggl(\boldsymbol{\mathcal{U}_0}+\boldsymbol{\epsilon}\biggr) &= \frac{q}{m}\mathbf{E} + \frac{q}{m}\biggl( \boldsymbol{\mathcal{U}_0} + \boldsymbol{\epsilon}\biggr) \times\mathbf{B}\nonumber \\
    \Rightarrow \dot{\boldsymbol{\epsilon}} &= \frac{q}{m}\boldsymbol{\epsilon} \times\mathbf{B}.
\end{align}
The solution to this equation exhibits oscillations at the gyrofrequency. As a result, explicit integration schemes must use timesteps smaller than the gyroperiod to resolve the instability, defeating the purpose of adopting a guiding center formalism in the first place.

\subsection{Semi-Implicit Numerical Integration Scheme}
\label{subsec: 2nd order semi-imp}

To resolve the numerical instability discussed in the previous subsection, we have formulated a second-order semi-implicit integration scheme that handles the numerical instability. Implicit schemes are well known for their stabilizing properties but often require numerical solutions of algebraic equations to take forward steps, leading to longer computation times compared to explicit schemes. Our approach combines the advantages of both methods by employing an implicit scheme only for the terms that are unstable, while using explicit integration for the remaining terms. Furthermore, because the implicit terms are linear and can, therefore, be solved analytically, the implicit aspect of the scheme does not require numerical solutions. We present the formal derivation of this second-order semi-implicit scheme in Appendix~\ref{appendix: 2nd order semi-implicit derivation}.

The scheme achieves second-order accuracy by following the second-order Runge-Kutta approach of first taking a half-step and then using the solution from the half-step to take a full step. The equations for the half-step from $\tau_n$ to $\tau_{n+1/2}$ is then
\begin{align}
\label{eqn:half-step}
    \frac{\mathcal{U}^\alpha_{n+1/2}-\mathcal{U}^\alpha_n}{\Delta\tau/2}=&
    \frac{q}{m}F^\alpha_{~\beta,n}\frac{\mathcal{U}^\beta_{n+1/2}+\mathcal{U}^\beta_n}{2} \\ \nonumber
    &\qquad-\Gamma^\alpha_{\beta\nu,n}\mathcal{U}^\beta_n \mathcal{U}^\nu_n
    -\mu_n\nabla^\alpha\omega_n, \\
    \chi^\alpha_{n+1/2} =& \chi^\alpha_n + \frac{\Delta\tau}{2} \mathcal{U}^\alpha_n,
\end{align}
where $\Delta \tau = \tau_{n+1} - \tau_n$. The full step from $\tau_n$ to $\tau_{n+1}$ is
\begin{align}
\label{eqn:full-step}
    \frac{\mathcal{U}^\alpha_{n+1}-\mathcal{U}^\alpha_n}{\Delta\tau}=&
    \frac{q}{m}F^\alpha_{~\beta,n+1/2}\frac{\mathcal{U}^\beta_{n+1}+\mathcal{U}^\beta_n}{2} \\ \nonumber
    &\qquad-\Gamma^\alpha_{\beta\nu,n+1/2}\mathcal{U}^\beta_{n+1/2} \mathcal{U}^\nu_{n+1/2}\nonumber\\
    & \qquad-\mu_{n+1/2}\nabla^\alpha\omega_{n+1/2}, \\
    \chi^\alpha_{n+1} =& \chi^\alpha_n + \Delta\tau \mathcal{U}^\alpha_{n+1/2}.
\end{align}
Note that both equations are implicit but linear and, therefore, can be easily evaluated without numerical solutions. Moreover, we do not employ these equations to step forward the $\mathcal{U}^t$ of the guiding center velocity. Instead, at every half step and full step, we evaluate this component using equation \eqref{eqn:conserved quantity}.

Formally, in the guiding center approximation, the magnetic moment is held constant. However, for reasons we will explain in \S\ref{subsec:magnetic moment}, the above integration scheme allows for the magnetic moment to be evolved.  Whether the magnetic moment is held constant or not, it does not affect the order of accuracy of the integration scheme. 

When numerically integrating the trajectory of a charged particle, the stepsize is restricted to be strictly less than the gyroperiod. In the case of the guiding center approach, this restriction is significantly relaxed: the stepsize must be smaller then the time it takes for the particle to drift into a region where the orientation and/or magnitude of the electromagnetic field are significantly different. To achieve this, we choose (see also eq.~\ref{eqn: gc assumption 2}) 
\begin{align}
\label{eqn: stepsize}
    \Delta\tau = \xi \frac{\omega}{\text{Max}\Bigl|\mathcal{U}^\nu\partial_\nu\frac{q}{m} F^\alpha_{~\beta}\Bigr|}\;.
\end{align}
We found that choosing $\xi=10^{-3}$ leads to results not limited by truncation error in the integration scheme, while not increasing significantly the computational cost.

\section{Integration with GRMHD Backgrounds}
\label{sec:GRMHD background}

In this section, we discuss the additional steps required to integrate the numerical method introduced in the previous section with the electromagnetic fields obtained from GRMHD simulations. Because the electromagnetic field is digitized on the GRMHD grid, this requires an interpolation method to reconstruct the field at the instantaneous position of the guiding center. In particular, because of magnetic mirroring, which is caused by field convergence, the method needs to generate smooth and accurate first derivatives to ensure reliable integration of particle trajectories. The same conclusion can be reached by examining the guiding center equation~\eqref{eqn:GC EOM}, which depends explicitly on the first derivatives of the electromagnetic field. To achieve this, we have implemented a tricubic interpolation scheme, ensuring that the first derivatives are continuous and allowing for the electromagnetic field to transition smoothly between cells or grid points. 

An artifact of interpolation methods is that they do not force the reconstructed electromagnetic field to remain a solution to the homogeneous Maxwell equations between the grid points. As we will show in \S\ref{subsec:magnetic moment}, this causes the magnetic moment of each charge to be not conserved at the appropriate order along the trajectory (\citealt{Ripperda2018AIntegrators} also found that the magnetic moment is not conserved when integrating particles on a discretized background versus on equivalent analytic ones). The magnitude of this artifact depends on the grid spacing, which we will explore in \S\ref{subsec:vary grid}.

Additionally, the interpolation method does not preserve the fact that the electric and magnetic fields in an ideal GRMHD flow are mutually perpendicular. We, therefore, run the risk of introducing parallel/non-ideal electric field components, which will artificially accelerate or decelerate the particles. There is an additional subtlety in this calculation, since the electromagnetic field tensor used in the covariant guiding center equation of motion is calculated at each step from the raw magnetic field and fluid velocity quantities produced by the GRMHD algorithm, in this case \texttt{Athena++} \citep{Stone2020TheSolvers}. We found that by first interpolating the magnetic field and fluid velocity from the GRMHD outputs to the position of the guiding center and then constructing the electromagnetic field at that point from the interpolated quantities, forcing the force-free condition, we avoid introducing non-ideal electric fields and related artifacts.

\subsection{Tricubic Interpolation}
\label{subsec:tricubic interp}

To evolve the positions of charges on a discretized electromagnetic field, such as that obtained from a GRMHD simulation, we need to interpolate the values of the magnetic field and fluid velocity at arbitrary positions. Here we chose a tricubic interpolation scheme derived in \cite{Lekien2005TricubicDimensions}, which takes the form
\begin{equation}
\label{eqn: tricubic form}
    f(x^\prime,y^\prime,z^\prime) = \sum_{i,j,k=0}^3 a_{ijk}\, x^{\prime} y^{\prime} z^{\prime}\;.
\end{equation}
Here, $f$ is the quantity to be interpolated and $x^\prime$, $y^\prime$, and $z^\prime$ are the appropriately scaled position coordinates such that
\begin{equation}
    x^\prime = \frac{x - x_i}{x_{i+1}-x_i},
\end{equation}
where $x_i$ is the nearest grid coordinate less than $x$. Note that, even though the position coordinates are written in $x$, $y$, and $z$ notation, this interpolation method works for non-Cartesian coordinate grids as well. 

The coefficients $a_{ijk}$ are obtained from the discretized values of the quantity $f$ such that the values of $f$, and of its first 3 derivatives, $df/dx$, $df/dy$, $df/dz$, $d^2f/dxdy$, $d^2f/dxdz$, $d^2f/dydz$, and $d^3f/dxdydz$ at the grid points agree with those evaluated using finite differences, i.e., 
\begin{equation}
    \frac{df(x^\prime_l,y^\prime_m,z^\prime_n)}{dx^\prime} = \frac{1}{2}\biggl[f(x^\prime_{l+1},y^\prime_m,z^\prime_n) - f(x^\prime_{l-1},y^\prime_m,z^\prime_n)\biggr],
\end{equation}
\begin{multline}
    \frac{d^2f(x^\prime_l,y^\prime_m,z^\prime_n)}{dx^\prime dy^\prime}= \\
    \frac{1}{4}\biggl[f(x^\prime_{l+1},y^\prime_{m+1},z^\prime_n)-f(x^\prime_{l-1},y^\prime_{m+1},z^\prime_n) \\ 
     - f(x^\prime_{l+1},y^\prime_{m-1},z^\prime_n)+ f(x^\prime_{l-1},y^\prime_{m-1},z^\prime_n) \biggr],
\end{multline}
\begin{multline}
    \frac{d^3f(x^\prime_l,y^\prime_m,z^\prime_n)}{dx^\prime dy^\prime dz^\prime}= \\
    \frac{1}{8}\biggl[f(x^\prime_{l+1},y^\prime_{m+1},z^\prime_{n+1}) + f(x^\prime_{l+1},y^\prime_{m-1},z^\prime_{n-1})\quad\\
    +f(x^\prime_{l-1},y^\prime_{m+1},z^\prime_{n-1}) + f(x^\prime_{l-1},y^\prime_{m-1},z^\prime_{n+1})\quad\quad\\
    -f(x^\prime_{l-1},y^\prime_{m-1},z^\prime_{n-1}) - f(x^\prime_{l+1},y^\prime_{m+1},z^\prime_{n-1})\quad\quad\\
    -f(x^\prime_{l+1},y^\prime_{m-1},z^\prime_{n+1}) - f(x^\prime_{l-1},y^\prime_{m+1},z^\prime_{n+1})\biggr]\;.
\end{multline}
Here $l$, $m$, and $n$ are coordinate grid indices and the derivatives with respect to the other coordinates follow similarly.

Conveniently, the tricubic form, as described in equation \eqref{eqn: tricubic form} allows for straightforward interpolation of the derivatives of $f$ by taking the derivatives of the tricubic form itself, e.g.,
\begin{equation}
    \label{eqn: derivative interpolation}
    \frac{df(x^\prime,y^\prime,z^\prime)}{dx^\prime} = \sum_{i=1,j,k=0}^3 a_{ijk}\,i\, x^\prime_{i-1} y^\prime_j z^\prime_k,
\end{equation}
The derivatives with respect to the other coordinates follow similarly. We employ this method to evaluate the derivatives of the gyrofrequency in the guiding center equation of motion, equation \eqref{eqn:GC EOM}. 

\subsection{Numerical Invariance of the Magnetic Moment}
\label{subsec:magnetic moment}

Formally, the magnetic moment is an adiabatic invariant for the settings we are considering \citep{Cary2009HamiltonianMotion, Burby2020GeneralSystems}. However, numerical, truncation, and round-off errors cause the magnetic moment of a charged particle to evolve with time.

\cite{Brizard2022OnField} showed that, for a non-uniform but constant-in-time magnetic field, the magnetic moment of a charged particles evolves at a rate proportional to the divergence of the magnetic field, which formally should be equal to zero. When we include the time dependence of the magnetic field as well as the presence of a perpendicular electric field, we find that, to the order of our approximations,
\begin{align}
\label{eqn: not covariant magnetic moment evolution}
    \biggl\langle \frac{d \mu}{dt} \biggr\rangle = &-\frac{\mu \,v_\|}{\sqrt{B^2-E^2}}\nabla \cdot \mathbf{B} \\ \nonumber
    &-\frac{\mu}{\sqrt{B^2-E^2}} \biggl(\frac{\partial \mathbf{B}}{\partial t}+ \nabla \times \mathbf{E}  \biggr)\cdot \mathbf{b}\;,
\end{align}
where $v_\|$ is the velocity of the particle parallel to the magnetic field, and $\mathbf{b}$ is the unit vector in the direction of the magnetic field (see Appendix~\ref{appendix: magnetic moment evolution}).

Equation \eqref{eqn: not covariant magnetic moment evolution} shows that for fields that are solutions to the homogeneous Maxwell equations, the zeroth-order magnetic moment is constant. However, this may not be the case for numerically evaluated electromagnetic fields, which obey Maxwell's equations up to a truncation (and perhaps interpolation) error.

To demonstrate the presence of these artifacts and quantify their magnitudes, we will keep track of the (non-physical) evolution of the magnetic moment of the guiding center. We first recast equation \eqref{eqn: not covariant magnetic moment evolution} into a covariant form by writing its 3+1 form for general spacetimes as
\begin{align}
\label{eqn: 3+1 magnetic moment evolution}
     \biggl\langle \frac{d \mu}{d\tau} \biggr\rangle =&
     -\frac{\mu \, \mathcal{U}_\|}{\sqrt{B^2-E^2}\sqrt{-g}}\partial_i (\sqrt{-g}B^i) \\ \nonumber
     &- \frac{\mu \, \mathcal{U}^t}{\sqrt{B^2-E^2}\sqrt{-g}}( \partial_t B^i + \epsilon^{0ijk}\partial_j E_k )b_i,
\end{align}
where the $i$, $j$, $k$ indices range from 1 to 3, and $\epsilon^{\lambda\nu\alpha\beta}=(-g)^{-1/2}[\lambda\nu\alpha\beta]$ \citep{Poisson2004AMechanics}. We now convert this to a covariant form using the definitions of the electric and magnetic fields in the coordinate frame in terms of the electromagnetic field tensor \citep{Komissarov20113+1Magnetodynamics}
\begin{align}
\label{eqn: covariant electric field tensor}
    E_i =& \,F_{t\,i}, \\ 
\label{eqn: covariant magnetic field tensor}
    B^i =& \,\frac{1}{2}\epsilon^{0ijk}F_{jk}.
\end{align}

Substituting the above into equation \eqref{eqn: 3+1 magnetic moment evolution}, we obtain 
\begin{align}
\label{eqn: covariant magnetic moment evolution 2}
    \biggl\langle \frac{d \mu}{d\tau} \biggr\rangle = -\frac{\mu}{\sqrt{2F_{\iota \eta}F^{\iota \eta}}} \zeta_\lambda \, \epsilon^{\lambda \nu\alpha\beta}\, \partial_\nu F_{\alpha\beta}\;,
\end{align}
where $\zeta_\lambda \equiv ( \mathcal{U}_\| , \mathbf{b})$. 

\subsection{Tracer Particles in a GRMHD Background}
\label{subsec: guiding center in GRMHD}

We numerically solve the guiding center equations of motion, equation \eqref{eqn:GC EOM}, using the semi-implicit integration scheme detailed in subsection~\ref{subsec: 2nd order semi-imp}, with a static GRMHD background. The GRMHD background was generated with \texttt{Athena++} \citep{Stone2020TheSolvers} and corresponds to a geometrically thick, standard and normal evolution (SANE) accretion flow around a black hole with a spin of $a=0.5$. We chose a snapshot of the electromagnetic field of the GRMHD simulation at $t=130,000\,$M, when the accretion flow has settled into a quasi-steady turbulent state. The effective grid resolution of the GRMHD simulation was 768 cells in the radial up to a distance of 1,000\,M, 288 cells in the poloidal direction, and 128 cells in the azimuthal direction. The details of this GRMHD simulation setup and analysis is presented in \citet[][in preparation]{Avara2025}.

\begin{figure*}
    \centering
    \gridline{\fig{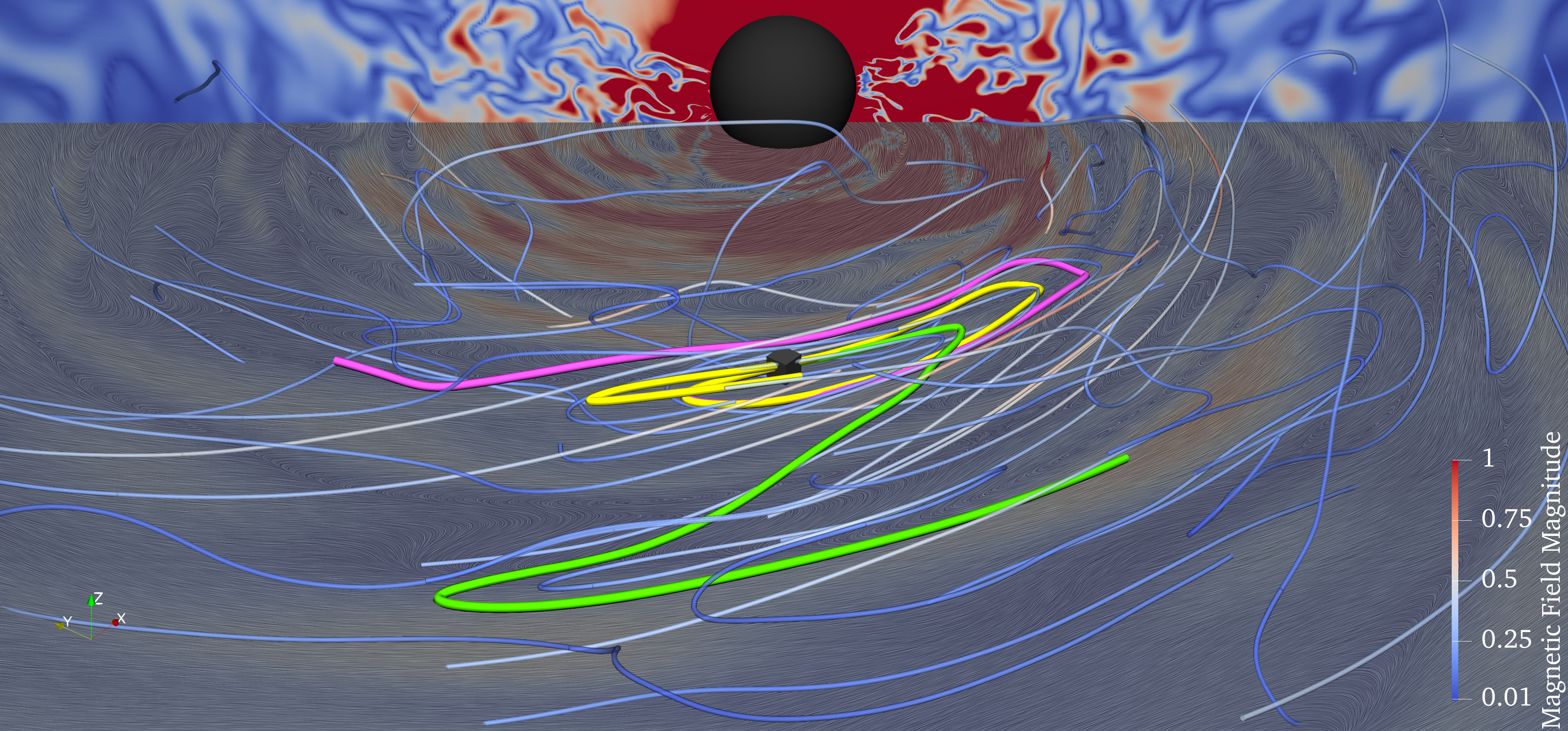}{\textwidth}{}}
    \gridline{\fig{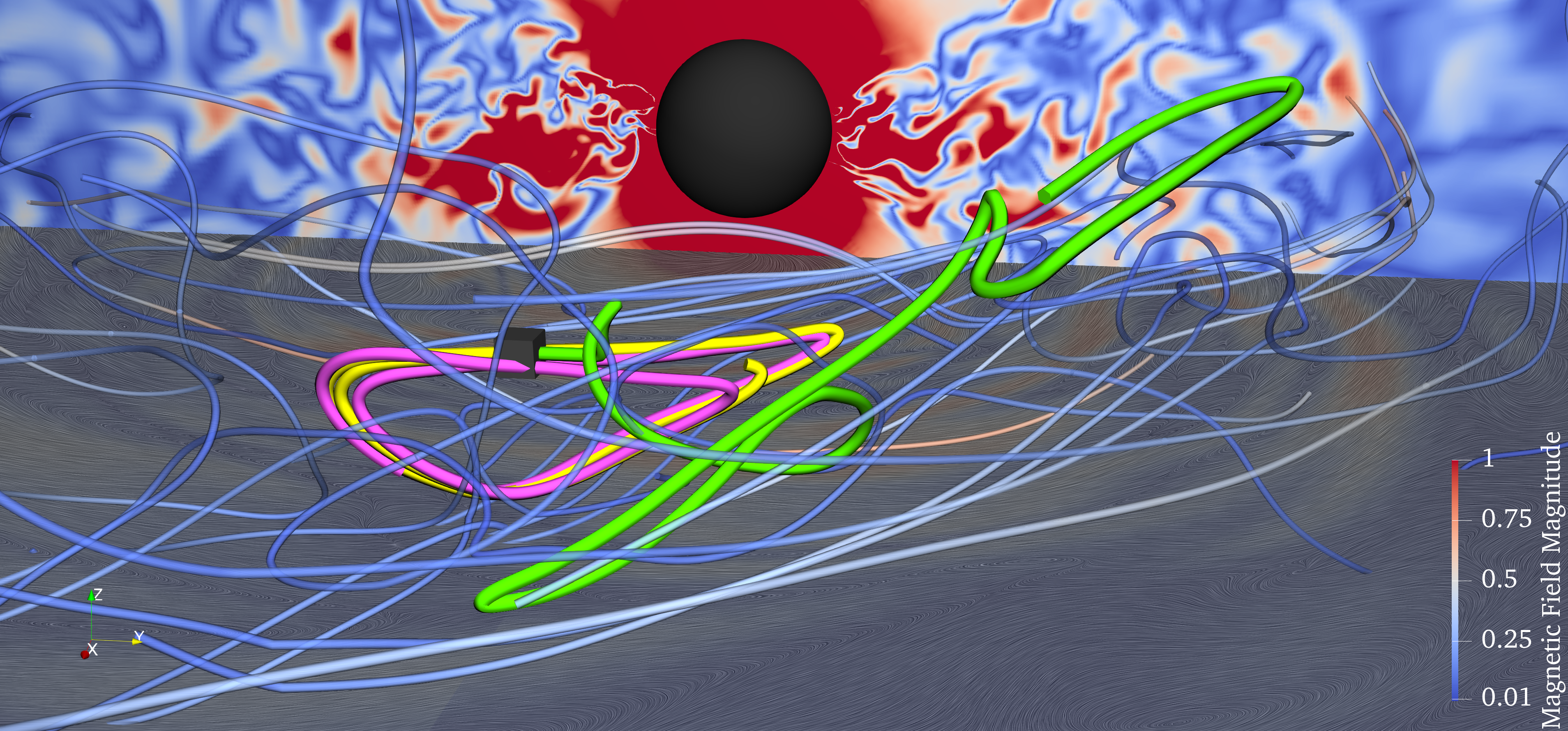}{\textwidth}{}}
    \caption{Guiding center trajectories in a GRMHD background, initiated at the locations of the black boxes, with particle velocities that are parallel (green), perpendicular (yellow), and anti-parallel (purple) to the local magnetic fields. The black sphere represents the black hole event horizon. The vertical and horizontal cross-sections show contours of magnetic field strength. The horizontal cross-section also includes projected magnetic field lines. A small selection of magnetic field lines near the guiding center trajectories are also shown with their color determined by the magnetic field strength.}
    \label{fig: grmhd viz}
\end{figure*}

We show the resulting guiding center trajectories  in Figure~\ref{fig: grmhd viz}. In the top panel, we initialized three guiding center trajectories in the equatorial plane at a radius of 12.8\,M, with initial velocities that are parallel, perpendicular, and anti-parallel to the local magnetic field. Similarly, in the bottom panel, we initialized three different guiding center trajectories in the equatorial plane at a radius of about 15\,M and at a different azimuth location around the black hole. We specifically selected these regions of the accretion flow because of the complexity in the local magnetic field line structures, making them ideal for evaluating the accuracy and effectiveness of our guiding center methodology.  In contrast, other regions in the flow that feature laminar, azimuthally-aligned magnetic field lines lead to simpler guiding center trajectories that trivially follow the large-scale field lines.

We initialized the trajectories with a Lorentz factor $\gamma=10$ and integrated them up to an elapsed coordinate time of 30\,M. We scaled the background electromagnetic field such that the initial gyroradii of the trajectories with particle velocities perpendicular to the local field were equal to $10^{-5}\,$M; the corresponding gyroradii for the other trajectories were $\simeq 10^{-6}\,$M. In the next section, we will use these trajectories to evaluate the performance of the numerical algorithm.

\section{Algorithmic Performance}
\label{sec: numerical characteristics}

Applying the guiding center approximation to calculate charged particle trajectories in black hole accretion flows obtained using GRMHD simulations involves several layers of approximations. These are: approximating the particle equation of motion by the equations for its guiding center, approximating the discretized electromagnetic field at arbitrary locations in the spacetime through interpolation, and using a numerical integration method for solving the differential equations of motion. In this section, we explore the impact of these approximations on the trajectories on particles, using as fiducial cases the trajectories with perpendicular initial pitch angle discussed in the previous section. 

\begin{figure*}
    \centering
    \gridline{\fig{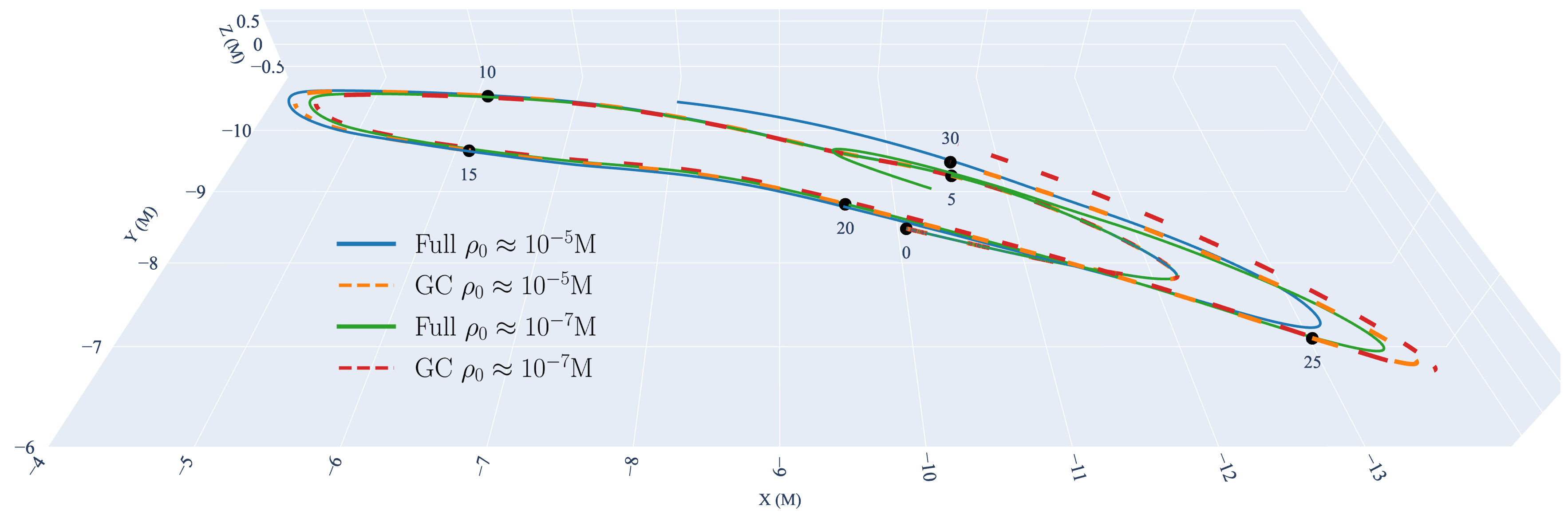}{\textwidth}{}}
    \caption{Comparison of particle trajectories calculated by integrating the guiding-center equations (labeled as ``GC'') and the full equations of motion for charged particles (label as ``Full''), for different values of the initial gyroradius $\rho_0$. Trajectories where initialized at a radius of $12.8M$, with an initial velocity orientation perpendicular to the local magnetic field and particle Lorentz factors of 10 (see also Fig.~\ref{fig: grmhd viz}). Black dots mark the elapsed coordinate time, in units of M.}
    \label{fig: traj comparison 1}
\end{figure*}

\begin{figure*}
    \centering
    \gridline{\fig{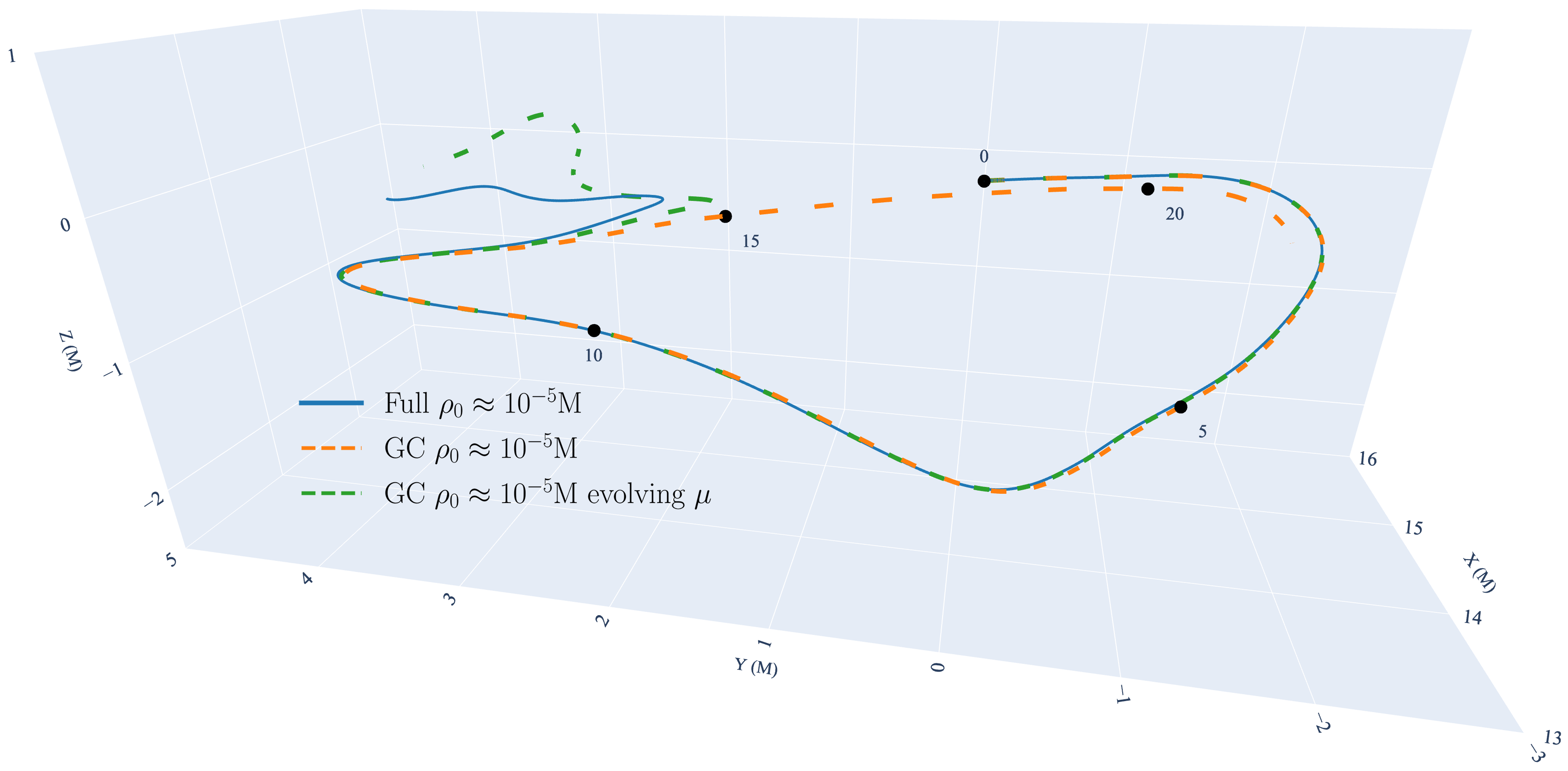}{\textwidth}{}}
    \caption{Same as Figure~\ref{fig: traj comparison 1} but for an initial radius of $15M$ and a different initial azimuthal location. Note that, in this case, the trajectories diverge at a coordinate time of $\sim 15M$. The dashed green curve depicts the guiding center trajectory for $\rho_0\simeq 10^{-5}M$  calculated for an evolving magnetic moment. }
    \label{fig: traj comparison 2}
\end{figure*}

To achieve this, we compare the guiding center trajectories with those calculated by integrating the full equations of motions of the charged particles with the same initial conditions. We solve the latter using a fourth-order Runge-Kutta (RK4) scheme with an adaptive stepsize chosen to be equal to $10^{-3}$ of the instantaneous gyroperiod.

Figure~\ref{fig: traj comparison 1} compares the full and guiding center trajectories for particles at 12.8\,M and with initial gyroradii of $10^{-5}\,$M and $10^{-7}\,$M, to assess the accuracy of the guiding center approximation as a function of magnetic field strength and, hence, gyroradius size. Surprisingly, the guiding center trajectory more closely matches the full trajectory for the larger initial gyroradius ($10^{-5}\,$M) than for the smaller gyroradius ($10^{-7}$\,). This result is counterintuitive, as a smaller gyroradius is expected to improve the validity of the guiding center approximation. Before explaining the origin of this discrepancy, however, we will first discuss the second set of fiducial trajectories.

Figure~\ref{fig: traj comparison 2} compares the full and guiding center trajectories for particles initialized at 15\,M and at different azimuthal location around the black hole. In this case, the full and guiding center trajectories diverge after a coordinate time of $\sim 15\,$M. 

\begin{figure}
    \centering
    \includegraphics[width=\linewidth]{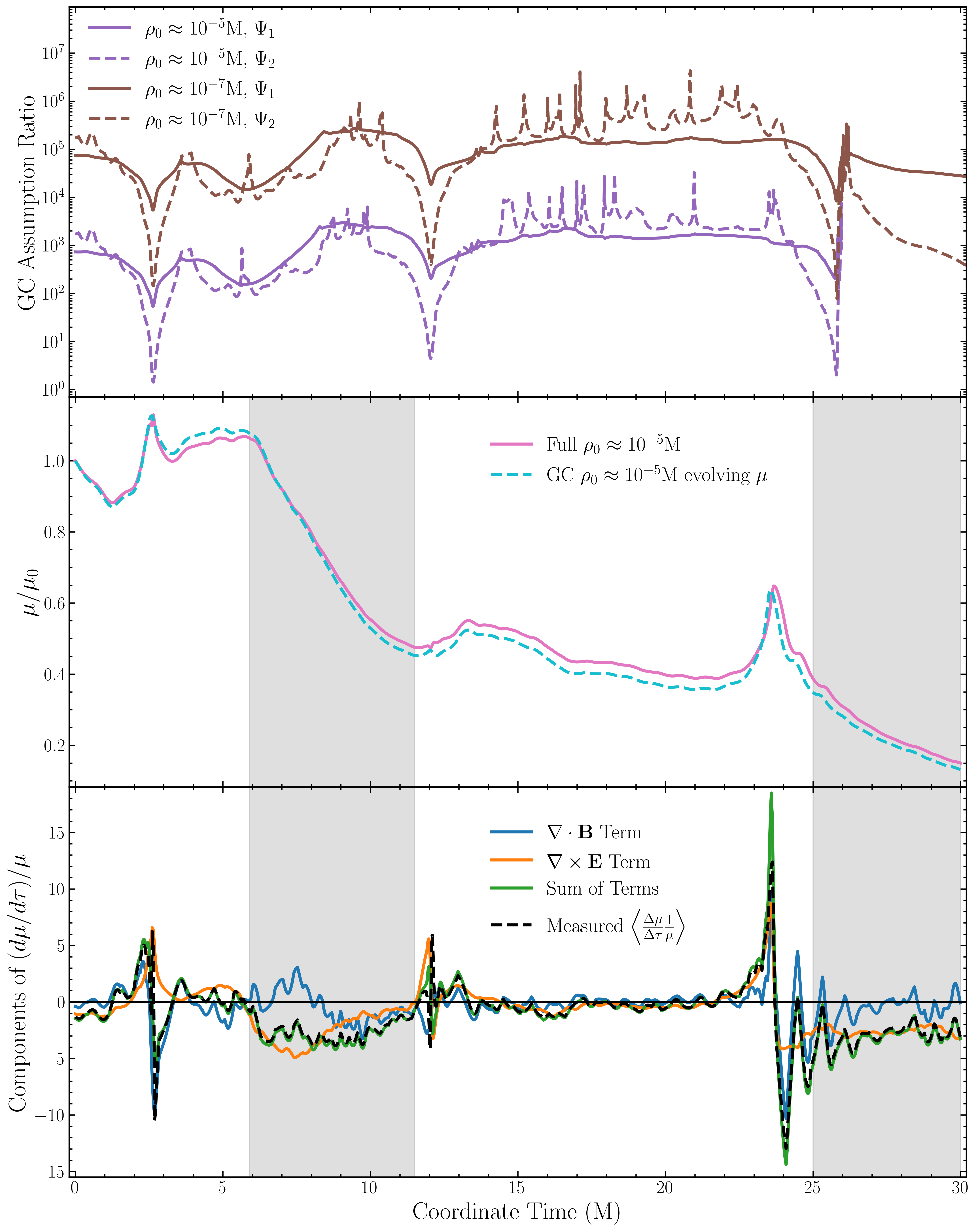}
    \caption{Characteristics of the trajectories shown in Fig.~\ref{fig: traj comparison 1}.
    \textit{(Top)\/} The ratios $\Psi_1$ and $\Psi_2$ that quantify the validity of the guiding center assumptions. 
    \textit{(Middle)\/} The magnetic moment of the charged particle, normalized to its initial value. For the full trajectory (purple), the magnetic moment is calculated at each timestep using equation \eqref{eqn: magnetic moment} for the local fields and instantaneous velocity. For the guiding center trajectory (dashed blue), the magnetic moment is calculated by evolving equation~\eqref{eqn: covariant magnetic moment evolution 2}.
    \textit{(Bottom)\/} Dashed black curve: the gyro-averaged rate of change of the magnetic moment normalized by the instantaneous magnetic moment as a function of coordinate time for the full trajectory. Blue curve: the divergence of the magnetic field caused by the interpolation scheme. Orange curve: the curl of the electric field. Green curve: the sum of the last two terms, demonstrating that they are the origin of the evolution of the magnetic moment.}
    \label{fig: r12 data}
\end{figure}

\begin{figure}
    \centering
    \includegraphics[width=\linewidth]{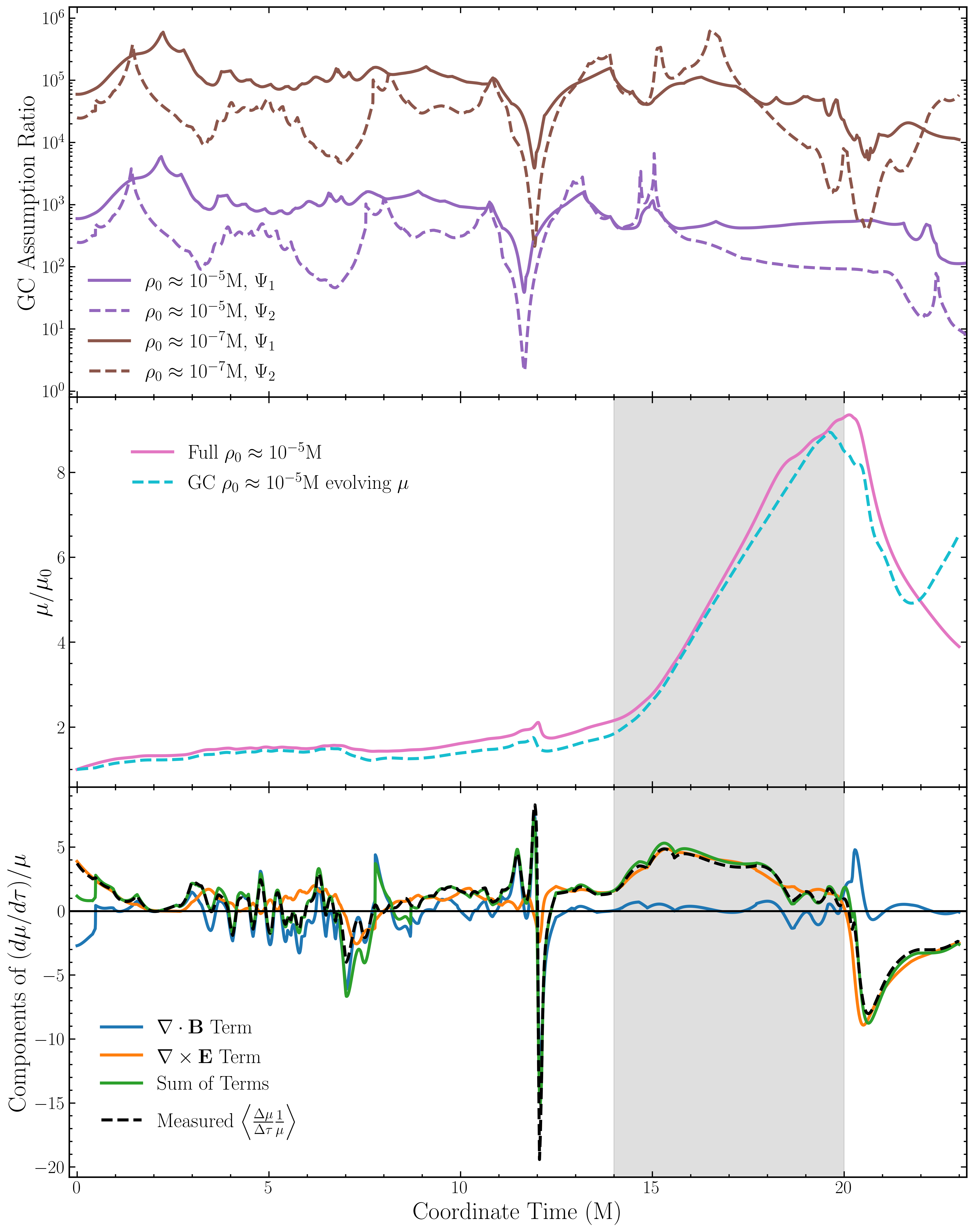}
    \caption{Same as Fig.~\ref{fig: r12 data} but for the trajectories shown in Fig.~\ref{fig: traj comparison 2}.}
    \label{fig: r15 data}
\end{figure}

In order to assess which of the underlying assumptions leads to this divergence, we show in the top panels of Figures~\ref{fig: r12 data} and \ref{fig: r15 data} the two ratios $\Psi_1$ and $\Psi_2$ that quantify the validity of the guiding center assumptions (see eqs.~\ref{eqn: gc assumption 1} and \ref{eqn: gc assumption 2}). In both cases, the large values of the ratios generally indicate that the guiding center assumptions hold. The occasional dips in these ratios occur at locations with large derivatives of the magnetic-field strength but appear to be only mildly correlated with the observed discrepancies.

The middle panels of Figures~\ref{fig: r12 data} and \ref{fig: r15 data} show the magnetic moment of the particle, as a function of coordinate time, normalized by the initial magnetic moment, for both full and guiding center trajectories. The magnetic moment of the full trajectory was calculated at every timestep using equation~\eqref{eqn: magnetic moment}. The magnetic moment for the guiding center trajectory was calculated by evolving equation~\eqref{eqn: covariant magnetic moment evolution 2}. Clearly, the large divergence between the trajectories at $t=15\,$M in Figure~\ref{fig: traj comparison 2} corresponds to a significant increase in the magnetic moment of the charge. The similarity between the evolution of the magnetic moments in both cases calculated with the full trajectories and those using equation~\eqref{eqn: magnetic moment} strongly suggests that this is simply an artifact introduced by the fact that the electromagnetic fields in the simulation do not obey the homogeneous Maxwell equations. 

To further demonstrate this point, the bottom panel of Figures \ref{fig: r12 data} and \ref{fig: r15 data} shows the gyro-averaged derivative of the magnetic moment with respect to proper time scaled by the current magnetic moment as a function of coordinate time for the full trajectory with initial gyroradius $10^{-5}$M. It also compares this with the magnitudes of the two terms in equation~\eqref{eqn: 3+1 magnetic moment evolution} that describe the evolution of the magnetic moment caused by the divergence of the magnetic field and the curl in the electric field (both normalized to the magnetic moment) as well as to their sum. Even though the discretization of the magnetic field introduces some degree of non-zero divergences, the evolution of the magnetic moment occurs at times when there is significant curl of the electric field. This is, however, an artifact of our simulation that uses a single GRMHD snapshot assuming that the background electromagnetic field is effectively static and, hence, that the time derivative of the magnetic field is zero. In a more realistic simulation, in which we will be evolving the GRMHD background, the time derivative of the magnetic field will counteract the effects of the curl of the electric field (see eq.~[\ref{eqn: 3+1 magnetic moment evolution}]), resulting in improved conservation of the magnetic moment.

As a final point of demonstration, Figure~\ref{fig: traj comparison 2} also includes a guiding-center trajectory for $\rho_0\simeq 10^{-5}M$ for which we have allowed for the magnetic moment to evolve in time. This trajectory matches more closely the one calculated by integrating the full equation of motion. It is interesting (albeit obvious in hindsight) that integrating the full equation of motion of the charges suffers from the same artifact, even though one would not have realized this without evaluating the evolution of the magnetic moment. 

\subsection{Impact of Background Field Resolution}
\label{subsec:vary grid}

The bottom panels of Figures~\ref{fig: r12 data} and \ref{fig: r15 data} show a small yet non-negligible residual in the numerical evaluation of the divergence of the magnetic field. This is caused by evaluating derivatives using the interpolation scheme (see eq.~[\ref{eqn: derivative interpolation}]) and introduces numerical artifacts to the integration of the guiding center equation of motion that explicitly depends on the electromagnetic field derivatives.

Here, we investigate how interpolation errors in the electromagnetic field derivatives contribute to discrepancies between the full particle trajectories and the guiding center trajectories by performing integrations in flat spacetime with a magnetic dipole configuration. To ensure that any differences arise solely from the interpolation of the electromagnetic field, we employ the RK4 method with identical stepsize protocols for integrating both the full and guiding center equations of motion. The charged particles are initialized in the equatorial plane at a radius of 1 (arbitrary units), with a Lorentz factor of 2, and an initial velocity pitch angle of 45 degrees relative to the local magnetic field. We integrate the particle for an elapsed coordinate time of 4 (arbitrary units); this is approximately equivalent to one full bounce cycle, i.e., after the particle reflects twice because of magnetic mirroring force and then returns roughly to the equatorial plane. For this setup, we can use the analytical solution for the mirror points \citep{Rossi1970IntroductionSpace}:
\begin{equation}
\label{eqn: pitch angle to mirror angle expression}
    \sin^2{\alpha_i} = \frac{\sin^6{\theta_m}}{(3\cos^2{\theta_m}+1)^{1/2}}
\end{equation}
to find that the reflections occur at radius of $\sim0.85$, and poloidal angles of $\sim67$ degrees and $\sim113$ degrees. Here $\alpha_i$ is the initial pitch angle and $\theta_m$ is the poloidal angle of the mirroring points. The radial position of the mirroring point can be calculated using the magnetic dipole field line relation $r_m=R_0\sin^2{\theta_m}$, where $R_0$ is the radius of the field line at the equatorial plane.

We computed both the full and guiding center trajectories using the analytical expression for the magnetic dipole and also using a discretized (gridded) representation of the magnetic field. For the latter, we evaluated the analytic expression on a spatial grid and then reconstructed the field at arbitrary points using the tricubic interpolation method described in \S\ref{subsec:tricubic interp}. We then compared the final positions of the guiding center to those calculated using the full trajectory with the same background resolution. 

\begin{figure}
    \centering
    \includegraphics[width=\linewidth]{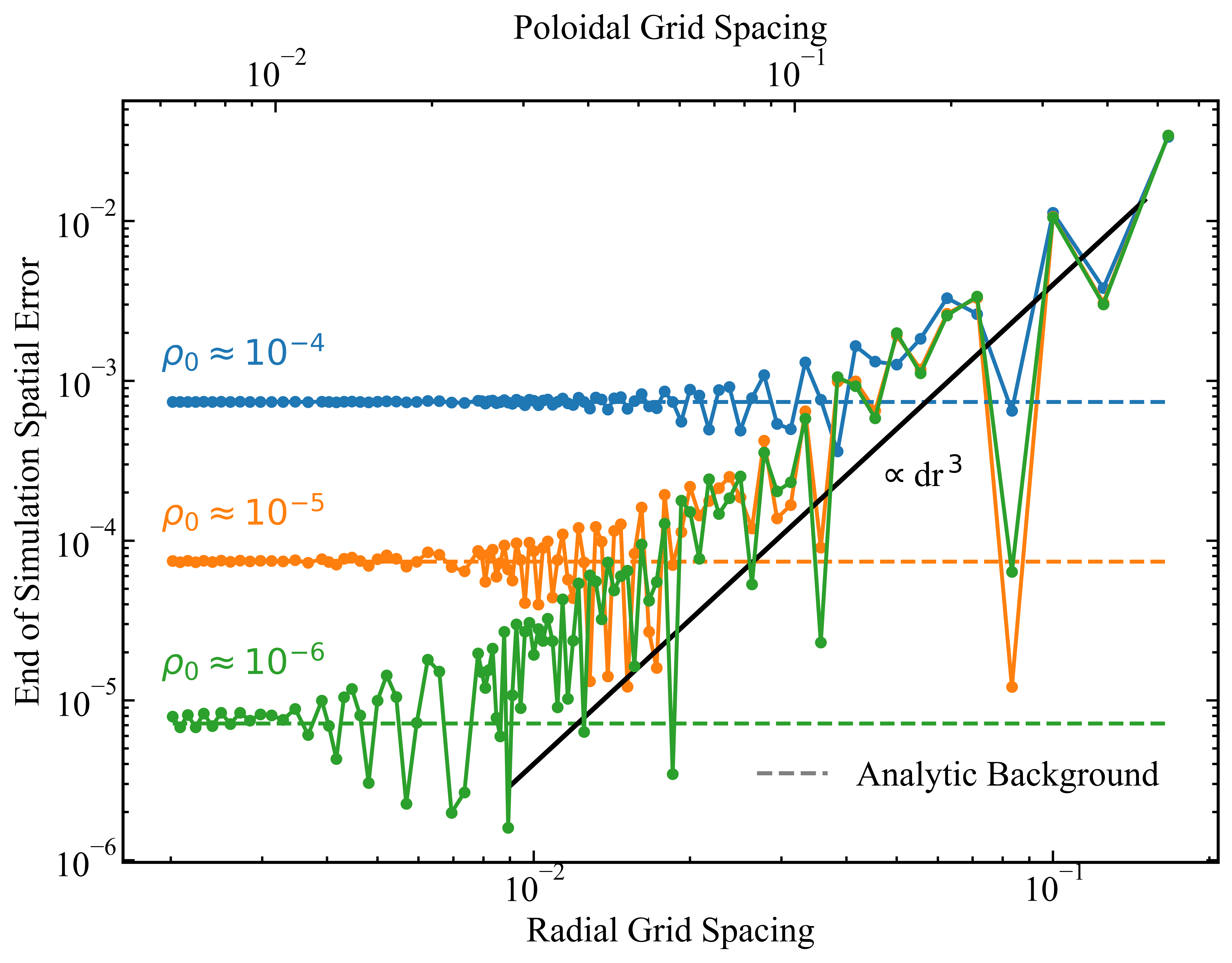}
    \caption{The difference in the end-of-simulation position between the guiding center and full charge trajectories as a function of grid spacing for the background electromagnetic field, for three different initial gyroradii $\rho_0$ (Solid curves). The horizontal dashed line shows the equivalent differences between the guiding center and full charge trajectories when the background field is evaluated analytically. The difference between the two approaches decreases proportionally to the cube of the grid spacing until reaching a floor imposed by the guiding center approximation.}
    \label{fig: grid error plot}
\end{figure}

Figure~\ref{fig: grid error plot} shows the difference in the final positions for simulations with different initial gyroradii and with different radial grid spacings; in each simulation the poloidal grid has twice as many points as the radial grid. The difference in the final positions between the two methods decreases proportionally to the cube of the grid spacing until reaching a floor. The cubic dependence of the error on the grid spacing matches the expected behavior of the third-order tricubic interpolation that we employ. To understand the origin of the floor, we repeated the exercise of comparing full trajectories to those of the guiding center but without interpolating the values of the electromagnetic field, as we know them everywhere analytically. For these runs, the differences in the final positions between the full and the guiding center trajectories are comparable to (but somewhat larger than) the size of the gyroradii, as expected given the approximations of the guiding-center equations. More importantly, these differences are equal to the floors identified in the previous comparisons. These results suggest that the error in the simulations scale with the grid spacing $\delta r$ as
\begin{equation}
    \epsilon \simeq \xi\rho_0+\delta r^3,
\end{equation}
with $\xi$ of order a few and all quantities are assumed to be in their natural units.

\section{Discussion \& Conclusions}
\label{sec:conclusion}

In this paper, we presented a numerical algorithm to integrate the covariant equations for the guiding center trajectories of charged particles in general spacetimes and in complex electromagnetic field geometries provided by GRMHD simulations. We find that the guiding center assumptions are, in general, well satisfied in the inner accretion flows, even in regions that are particularly turbulent, with complex magnetic field structures. 

The integration algorithm is semi-implicit in order to overcome a numerical instability inherent in the equations. It also employs a tricubic interpolation of the background fields to minimize artificial drift and mirroring effects introduced by non-continuous derivatives across grid cells. Importantly, we showed that any charged particle simulation that interpolates electromagnetic fields from discretized grids, whether employing guiding center approximations or not, is susceptible to artifacts arising from the interpolation of discretized electromagnetic fields. In particular, we demonstrated that such interpolation introduces violations of the homogeneous Maxwell equations, leading to artificial evolution of the magnetic moment and deviations in particle trajectories. 

We explored how the grid resolution of the background field constrains the ability to maintain a divergence-free magnetic field. This places a practical limit on grid resolution, for which we devised an approximate scaling. Moreover, using a single static GRMHD snapshot rather than evolving the background self-consistently fails to capture the full time-dependence of the electromagnetic fields. This leads to spurious contributions to the evolution of the magnetic moment, particularly from unbalanced curls of the electric field. 

The results of this work suggest a clear need for electromagnetic field interpolation schemes that preserve the homogeneous Maxwell equations. While interpolating the vector potential is one possible approach \citep{Kempski2023CosmicReversals}, it is often not readily available and computationally prohibitive to solve for in large GRMHD simulations. As a result, direct interpolation of the fields remains the practical alternative. In addressing grid instabilities in Particle-In-Cell simulations, studies such as \cite{Adams2025GridAlgorithms} and \cite{Werner2025SuppressingSmoothing} have explored alternative field interpolation and smoothing techniques. The severity of interpolation artifacts depends strongly on the underlying electromagnetic field geometry. However, the extent to which such artifacts may have biased prior hybrid particle-MHD studies—such as those on particle confinement in turbulent plasmas relevant to solar flares, cosmic rays, and black hole accretion flows—remains uncertain.

Future directions include applying the hybrid drift-kinetic GRMHD algorithm to time-dependent backgrounds. Such simulations have the potential to address long-standing questions in astrophysical particle transport. In particular, they could shed light on the flaring activity observed from Sgr~A*, where it remains unclear how ultra-relativistic particles—possibly produced in episodic magnetic reconnection events—remain confined for tens of minutes to over an hour, appearing as orbiting hot spots on the projected sky in the near-infrared \citep{Collaboration2023PolarimetryA}.

\begin{acknowledgements}

We thank Mark Avara for providing the black hole accretion flow simulation data and useful discussions and suggestions that contributed significantly to the manuscript. We thank Devin Bayly’s assistance with visualizations made in Paraview, which was made possible through University of Arizona Research Technologies visualization consulting service. This work has been supported by NSF PIRE award OISE-1743747. T.T. acknowledges support from the Alfred P. Sloan Foundation and the Ford Foundation. This material is based upon High Performance Computing resources supported by the University of Arizona TRIF, UITS, and Research, Innovation, and Impact and maintained by the UArizona Research Technologies department. At the Georgia Institute of Technology, this work used the Hive cluster, which is supported by the National Science Foundation under grant number 1828187, as well as research cyberinfrastructure resources and services provided by the Partnership for an Advanced Computing Environment (PACE). 
\end{acknowledgements}

\appendix

\section{Second-Order Semi-Implicit Integration Scheme Derivation}
\label{appendix: 2nd order semi-implicit derivation}

The covariant guiding center equations of motion (eq. \ref{eqn:GC EOM}) have a non-linear term in velocity, which is stable, and a linear term in velocity, which is unstable. To overcome this, we developed a second order semi-implicit integration scheme that we derive here. We show the derivation in one-dimension for brevity but it can be trivially generalized to multi-dimensions. 

Consider a set of 1D differential equations, with the acceleration equation having a linear and non-linear term, i.e.,
\begin{align}
    \label{eqn: 2nd semi-imp start dvdt}
    \frac{dv}{dt} &= \underbrace{f(x,v)}_{non-linear} +~\underbrace{g(x)~ v}_{linear},\\
    \label{eqn: 2nd semi-imp start dxdt}
    \frac{dx}{dt} &= v.
\end{align}
To achieve second-order accuracy in $\Delta t=h$, we will use a half way point in the integration, similar to that of the second order Runge-Kutta method.

First, we expand the velocity and position backwards from the half step, i.e., $n+1/2\rightarrow n$, as
\begin{gather}
    \label{eqn: v expanded back from half step}
    v_n = v_{n+1/2}-\frac{h}{2}\frac{dv}{dt}\biggr\vert_{n+1/2}+\frac{1}{2}\biggl(\frac{h}{2}\biggr)^2\frac{d^2v}{dt^2}\biggr\vert_{n+1/2}+\mathcal{O}(h^3),\\
    \label{eqn: x expanded back from half step}
    x_n = x_{n+1/2}-\frac{h}{2}\frac{dx}{dt}\biggr\vert_{n+1/2}+\frac{1}{2}\biggl(\frac{h}{2}\biggr)^2\frac{d^2x}{dt^2}\biggr\vert_{n+1/2}+\mathcal{O}(h^3).
\end{gather}
Next, we expand forward from the half step, i.e., $n+1/2\rightarrow n+1$:
\begin{gather}
    \label{eqn: v expanded forward from half step}
    v_{n+1} = v_{n+1/2}+\frac{h}{2}\frac{dv}{dt}\biggr\vert_{n+1/2}+\frac{1}{2}\biggl(\frac{h}{2}\biggr)^2\frac{d^2v}{dt^2}\biggr\vert_{n+1/2}+\mathcal{O}(h^3),\\
    \label{eqn: x expanded forward from half step}
    x_{n+1} = x_{n+1/2}+\frac{h}{2}\frac{dx}{dt}\biggr\vert_{n+1/2}+\frac{1}{2}\biggl(\frac{h}{2}\biggr)^2\frac{d^2x}{dt^2}\biggr\vert_{n+1/2}+\mathcal{O}(h^3).
\end{gather}
Subtracting equations~\eqref{eqn: v expanded back from half step} and \eqref{eqn: x expanded back from half step} from equations \eqref{eqn: v expanded forward from half step} and \eqref{eqn: x expanded forward from half step}, respectively, and solving for the $n+1$ step, we obtain
\begin{align}
    \label{eqn: error in v i+1}
    v_{n+1} =& v_n+h\frac{dv}{dt}\biggr\vert_{n+1/2} + \mathcal{O}(h^3),\\
    \label{eqn: error in x i+1}
    x_{n+1} =& x_n+h\frac{dx}{dt}\biggr\vert_{n+1/2} + \mathcal{O}(h^3).
\end{align}
The above shows that since the derivatives are already multiplied by a factor of $h$, as long as the derivatives are calculated to first order in $h$, then $v_{n+1}$ and $x_{n+1}$ will have second-order convergence. The derivatives in the above follow from equations \eqref{eqn: 2nd semi-imp start dvdt} and \eqref{eqn: 2nd semi-imp start dxdt}:
\begin{align}
    \label{eqn: half step dvdt}
    \frac{dv}{dt}\biggl\vert_{n+1/2} =& f(x_{n+1/2},v_{n+1/2})+g(x_{n+1/2})v_{n+1/2},\\
    \label{eqn: half step dxdt}
    \frac{dx}{dt}\biggl\vert_{n+1/2} =& v_{n+1/2}.
\end{align}
Substituting the above derivatives into equations \eqref{eqn: error in v i+1} and \eqref{eqn: error in x i+1} gives us the equations that evolve $v$ and $x$ from step $n$ to $n+1$
\begin{align}
    \label{eqn:evolve v}
    v_{n+1} =& v_n + h\, [ f(x_{n+1/2},v_{n+1/2})+g(x_{n+1/2})v_{n+1/2} ]+ \mathcal{O}(h^3),\\
    \label{eqn:evolve x}
    x_{n+1} =& x_n+h \, v_{n+1/2}+ \mathcal{O}(h^3).
\end{align}

We recall now that the linear term is unstable and, therefore, we would like to evaluate that term implicitly. To do so, we substitute $v_{n+1/2}$ for an implicit form. Starting from the addition of equations \eqref{eqn: v expanded back from half step} and \eqref{eqn: v expanded forward from half step}, we obtain
\begin{equation}
    \label{eqn: implicit v_n+1/2}
    v_n + v_{n+1} = 2v_{n+1/2} + \biggl(\frac{h}{2}\biggr)^2 \frac{d^2v}{dt^2}\biggr\vert_{n+1/2}+\mathcal{O}(h^3).
\end{equation}
Solving for $v_{n+1/2}$ and substituting that into equation \eqref{eqn:evolve v}, we find
\begin{equation}
    \label{eqn: v_n+1}
    v_{n+1} = v_n + h\,f(x_{n+1/2},v_{n+1/2})+h\,g(x_{n+1/2})\frac{v_n+v_{n+1}}{2} + \mathcal{O}(h^3),
\end{equation}
where we have dropped higher order terms in $h$. The above equation can be inverted analytically to solve for $v_{n+1}$. All that is left to do is solve for $x_{n+1/2}$ and $v_{n+1/2}$ to second-order accuracy in $h$, because the terms that depend on values at $n+1/2$ are already multiplied by a factor of $h$.

We again start from
\begin{equation}
    \label{eqn: v_n+1/2 forward expansion}
    v_{n+1/2} = v_n +\frac{h}{2}\frac{dv}{dt}\biggr\vert_n+\frac{1}{2}\biggl(\frac{h}{2}\biggr)^2\frac{d^2v}{dt^2}\biggr\vert_n+\mathcal{O}(h^3)\;.
\end{equation}
Subtracting equation \eqref{eqn: v expanded forward from half step} from the above, we find
\begin{equation}
    \label{eqn: v_n+1/2 -v_n (1)}
    v_{n+1/2}-v_n=v_n-v_{n_+1/2} +\frac{h}{2}\biggl(\frac{dv}{dt}\biggr\vert_n + \frac{dv}{dt}\biggr\vert_{n+1/2}  \biggr)
    +\frac{1}{2}\biggl(\frac{h}{2}\biggr)^2\biggl[\frac{d^2v}{dt^2}\biggr\vert_n-\frac{d^2v}{dt^2}\biggr\vert_{n+1/2} \biggr]+\mathcal{O}(h^3).
\end{equation}
The term in the square bracket is a difference of two derivatives evaluated at a distance $h/2$, so it is at least of first order in $h$. When this term is multiplied by $(h/2)^2$, it becomes of order $\mathcal{O}(h^3)$ and, thus, we can neglect it. Simplifying this expression gives
\begin{equation}
    \label{eqn: v_n+1/2 -v_n (2)}
    v_{n+1/2} -v_n = \frac{h}{4}\biggl(\frac{dv}{dt}\biggr\vert_n + \frac{dv}{dt}\biggr\vert_{n+1/2}\biggr)+\mathcal{O}(h^3).
\end{equation}

This last equation can now be inverted analytically by writing the derivatives as
\begin{align}
    \label{eqn:velocity derivative at n}
    \frac{dv}{dt}\biggr\vert_n &= f(x_n,v_n)+g(x_n)v_n,\\
    \label{eqn:velocity derivative at n+1/2}
    \frac{dv}{dt}\biggr\vert_{n+1/2} &= f(x_n,v_n)+g(x_n)v_{n+1/2}+\mathcal{O}(h).
\end{align}
Substituting the above derivatives into equation~\eqref{eqn: v_n+1/2 -v_n (2)}, we obtain
\begin{equation}
    \label{eqn: v_n+1/2 semi imp}
    v_{n+1/2}=v_n+\frac{h}{2}f(x_n,v_n) + \frac{h}{2}g(x_n)\frac{v_n+v_{n+1/2}}{2}+\mathcal{O}(h^2).
\end{equation}
which we can solve for $v_{n+1/2}$ analytically. To solve for $x_{n+1/2}$, we just take a forward half step, i.e.,
\begin{equation}
    \label{eqn: x_n+1/2}
    x_{n+1/2} = x_n +\frac{h}{2}v_n+\mathcal{O}(h^2).
\end{equation}

\section{Calculating the Electromagnetic Field Tensor from GRMHD Quantities}
\label{appendix: construct EM tensor}

In this appendix, we describe the procedure to couple the semi-implicit drift-kinetic integrator to the GRMHD simulation data output from \texttt{Athena++}. The guiding center equation of motion, equation \eqref{eqn:GC EOM}, is written in terms of the covariant electromagnetic field tensor but \texttt{Athena++} solves for the magnetic field in the 3+1 formalism. In order to use the fields solved for by the GRMHD simulations in the guiding center equations of motion, we need to construct the field tensor from the magnetic fields and fluid velocities that are output in the GRMHD simulation. Additionally, in our simulations we used Boyer-Lindquist coordinates, whereas \texttt{Athena++} uses Kerr-Schild spherical coordinates.

We start by constructing the velocity four-vector, $u^\mu$, and magnetic four-vector, $b^\mu$, from the three-vector quantities, $\Tilde{u}^i$ and $B^i$, output from \texttt{Athena++} as
\begin{align}
    u^t =& \frac{\gamma}{\alpha}, \\
    u^i =& \Tilde{u}^i - \frac{\gamma \beta^i}{\alpha}, \\
    b^t =& g_{ij}B^i u^j, \\
    b^i =& \frac{1}{u^t}(B^i+b^tu^i),
\end{align}
where index $\mu$ ranges over the space and time components, index $i$ and $j$ ranges over only the space components, $\gamma = \sqrt{1+g_{ij}\Tilde{u}^i\Tilde{u}^j}$, $\alpha=\sqrt{-1/g^{tt}}$, $\beta^i=g^{ti}\alpha^2$, and $g_{ij}$ is the Kerr-Schild metric in spherical coordinates \citep{Gammie2003HARM:Magnetohydrodynamics,McKinney2004AHole,Wong2022PATOKA:Accretion}.

Next we transform the velocity and magnetic four-vectors from Kerr-Schild spherical coordinates to Boyer-Lindquist coordinates. The $r$ and $\theta$ coordinates are the same in both coordinate systems and the transformation for the time and azimuthal angles are (see, e.g., \citealt{Christian2021FANTASY:Differentiation})
\begin{align}
\label{eqn: KS to BL t coord}
    t_{KS} = t_{BL} +\int \frac{2 r}{\Delta}dr,\\
\label{eqn: KS to BL r coord}
    \phi_{KS} = \phi_{BL} +\int \frac{a}{\Delta}dr,\\
\end{align}
where $a$ is the spin of the black hole, $\Delta = r^2 - 2 r +a^2$, and the subscripts BL and KS refer to the vector in Boyer-Lindquist or Kerr-Schild coordinates, respectively. The only four-vector components that transform between these coordinate systems are the time and azimuthal component \citep{McKinney2004AHole,Christian2021FANTASY:Differentiation}
\begin{align}
    u^t_{\text{BL}} =& u^t_{\text{KS}} - u^r_{\text{KS}} \frac{2 r}{\Delta}, \\
    u^\phi_{\text{BL}} =& u^\phi_{\text{KS}} - u^r_{\text{KS}} \frac{a}{\Delta}.
\end{align}
The magnetic four-vector transforms the same as the velocity four-vector. From this point on, we are working in Boyer-Lindquist coordinates and therefore drop the subscripts.

From the velocity and magnetic four-vector we construct the electromagnetic field tensor in the coordinate frame as \citep{Komissarov1999NumericalJets,Baumgarte2010NumericalComputer,Gammie2003HARM:Magnetohydrodynamics}
\begin{align}
    F_{\lambda\nu} = \epsilon_{\lambda\nu\alpha\beta}u^\alpha b^\beta,
\end{align}
where $\epsilon_{\lambda\nu\alpha\beta} = (-g)^{1/2}[\lambda\nu\alpha\beta]$ is the Levi-Civita tensor \citep{Misner1973Gravitation,Poisson2004AMechanics}, and $g$ is the determinant of the covariant Boyer-Lindquist metric.

\section{Derivation of Magnetic Moment Evolution Equation}
\label{appendix: magnetic moment evolution}

In this section we derive the evolution of the zeroth order gyro-averaged magnetic moment presented in equation~\eqref{eqn: not covariant magnetic moment evolution}. For this entire section we are only concerned with the zeroth order magnetic moment, which we will refer to as just the magnetic moment.

\cite{Brizard2022OnField} showed that, in the absence of an electric field, the change in the gyro-averaged magnetic moment is proportional to the divergence of the magnetic field, i.e., the magnetic moment is constant. Additionally, since the magnetic moment is a scalar and thus, a Lorentz invariant, the fact that it is constant in one frame, means also that it is constant in every frame. We extend here the \cite{Brizard2022OnField} derivation to include electric and magnetic fields where $\mathbf{E}\cdot\mathbf{B}=0$, i.e., the fields are perpendicular to each other, as is the case in GRMHD backgrounds. In this case, we can Lorentz transform the fields into a frame that has only a magnetic field and no electric field. Additionally, since there is no electric field in that frame, Faraday's law of induction,
\begin{equation}
    \label{eqn: faradays law}
    -\frac{\partial \mathbf{B}}{\partial t} = \nabla \times \mathbf{E},
\end{equation}
allows us to conclude that the magnetic field is constant in time. Therefore, in that frame, the evolution of the gyro-averaged magnetic moment reduces to what has been derived by \cite{Brizard2022OnField}, i.e.,
\begin{equation}
    \label{eqn: Brizard magnetic moment derivative}
    \biggl\langle \frac{d\mu}{dt}\biggr\rangle = -\frac{\mu\, v_\|}{B}\nabla\cdot\mathbf{B},
\end{equation}
where $v_\|$ is the parallel velocity component of the charged particle with respect to the magnetic field.

We now derive the equation for the evolution of the gyro-averaged magnetic moment for perpendicular magnetic and electric fields in a general frame. Our derivation follows closely the one by \cite{Brizard2022OnField}, except with the addition of a time-dependence of the magnetic field, and a term that is a function of the electric field, that is to be solved for.

We start by writing the simplified magnetic moment for the case with no electric fields, i.e.,
\begin{equation}
\label{eqn: Brizard magnetic moment}
    \mu = \frac{m v_\perp^2}{2 B},
\end{equation}
where $v_\perp$ is the perpendicular velocity component of the charged particle with respect to the magnetic field. The equation for the total time derivative of the magnetic moment becomes
\begin{align}
\label{eqn: derivative of mag mom 1}
    \frac{d \mu}{d t} = \frac{m\, \boldsymbol{v}_\perp\cdot \dot{\boldsymbol{v}}_\perp}{B} 
    - \frac{\mu}{B}\biggl(\frac{\partial B}{\partial t} + \boldsymbol{v}\cdot\nabla B\biggr)
    + \xi,
\end{align}
where we have added a placeholder term $\xi$ that will depend on the derivative of the electric field. By the end of the derivation, we will be able to solve for this term.

Next, we fully expand the first and third terms in the right-hand-side of this equation to obtain
\begin{align}
\label{eqn: derivative of mag mom 2}
    \frac{d \mu}{d t} = \frac{m}{B} \boldsymbol{v}_\perp\cdot
    \biggl[\dot{\boldsymbol{v}} 
    - \dot{v}_\| \mathbf{b} 
    - v_\|\biggl(\frac{\partial \mathbf{b}}{\partial t} + \boldsymbol{v}_\perp \cdot \nabla \mathbf{b} + v_\| \mathbf{b}\cdot\nabla\mathbf{b}\biggr)\biggr]
    - \frac{\mu}{B}\biggl(\frac{\partial B}{\partial t} + \boldsymbol{v}_\perp \cdot \nabla B + v_\| \mathbf{b}\cdot\nabla B \biggr)
    + \xi\;,
\end{align}
where $\mathbf{b}$ is a unit vector that is parallel to the magnetic field.

Substituting $\dot{\boldsymbol{v}} = q/m \boldsymbol{v}\times\boldsymbol{B}$ in the above equation and distributing the dot product with perpendicular velocity results in
\begin{align}
\label{eqn: derivative of mag mom 3}
    \frac{d \mu}{d t} =& - \frac{m v_\|}{B} \boldsymbol{v}_\perp \cdot (\boldsymbol{v}_\perp \cdot \nabla\mathbf{b})  
    - \frac{\mu\,v_\|}{B} \mathbf{b} \cdot \nabla B 
    -\frac{\mu}{B} \frac{\partial B}{\partial t}
    + \xi \\ \nonumber
    & - \frac{m v_\|}{B} \boldsymbol{v}_\perp \cdot \frac{\partial \mathbf{b}}{\partial t}
    - \frac{m v_\|^2}{B} \boldsymbol{v}_\perp \cdot (\mathbf{b}\cdot\nabla\mathbf{b})
    - \frac{\mu}{B} \boldsymbol{v}_\perp \cdot \nabla B.
\end{align}

The result of gyro-averaging the various terms in the above equation only depend on the factors of the perpendicular velocity, namely,
\begin{align}
\label{eqn: gyro-averaged perp velocity}
    \langle \boldsymbol{v}_\perp \rangle = & \,0 \\
\label{eqn: gyro-averaged perp velocity squarred}
    \langle \boldsymbol{v}_\perp \cdot (\boldsymbol{v}_\perp \cdot \nabla\mathbf{b}) \rangle = & 
    \,\frac{v_\perp^2}{2} \nabla \cdot \mathbf{b}.
\end{align}

Substituting the above into equation \eqref{eqn: derivative of mag mom 3} gives
\begin{align}
\label{eqn: derivative of mag mom 4}
    \biggl\langle \frac{d \mu}{d t}\biggr \rangle =&
    -\frac{\mu v_\|}{B}(B \nabla\cdot\mathbf{b} + \mathbf{b}\cdot\nabla B)
    - \frac{\mu}{B}\frac{\partial B}{\partial t} + \xi, \\
    &-\frac{\mu v_\|}{B}\nabla\cdot \mathbf{B}
    - \frac{\mu}{B}\frac{\partial \mathbf{B}}{\partial t} \cdot \mathbf{b}+ \xi.
\end{align}

We can now solve for the placeholder term $\xi$ recalling that, following Faraday's law of induction (eq.~[\ref{eqn: faradays law}]), the magnetic moment has to be invariant. For consistency, we also update the denominator in this expression, which arises from the expression of the gyrofrequency that also gets  modified in the presence of an electric field. The final result is then
\begin{align}
\tag{\ref{eqn: not covariant magnetic moment evolution}}
    \biggl\langle \frac{d \mu}{dt} \biggr\rangle = &-\frac{\mu \,v_\|}{\sqrt{B^2-E^2}}\nabla \cdot \mathbf{B} \\ \nonumber
    &-\frac{\mu}{\sqrt{B^2-E^2}} \biggl(\frac{\partial \mathbf{B}}{\partial t}+ \nabla \times \mathbf{E}  \biggr)\cdot \mathbf{b}.
\end{align}
Converting the above equation to a covariant expression is trivial since the terms in the equation are the homogeneous Maxwell equations which have a well-known covariant form.

\bibliography{references}
\bibliographystyle{aasjournal}
\end{document}